\documentclass[11pt]{article}
\usepackage{amsmath,amsthm,amscd,array,amssymb}

\begin{document}

\begin{center}
{\Large{\bf Euclidean super Yang--Mills theory
\\
\medskip\smallskip
on a hyper--K\"ahler eightfold}}
\\
\bigskip\medskip
{\large{\sc D. M\"ulsch}}$^{a}$ \footnote{Email:
muelsch@informatik.uni-leipzig.de}
and
{\large{\sc B. Geyer}}$^b$
\footnote{Email: geyer@itp.uni-leipzig.de}
\\
\smallskip
{\it$\!\!\!\!\!^a$ Wissenschaftszentrum Leipzig e.V., D--04103
Leipzig, Germany
\\\smallskip
 $^b$ Universit\"at Leipzig,
Naturwissenschaftlich-Theoretisches Zentrum\\ $~$ and Institut
f\"ur Theoretische Physik, D--04109 Leipzig, Germany}
\\
\bigskip
{\small{\bf Abstract}}
\\
\end{center}

\begin{quotation}
\noindent {\small{The construction of a $N_T = 3$ cohomological
gauge theory on the hyper--K\"ahler eight--fold, whose group
theoretical description was given previously by Blau and Thompson
\cite{1}, is performed explicitly.}}
\end{quotation}

\bigskip\medskip
%%%%%%%%%%%%%%%%%%%%%%%%%%%%%%%%%%%%%%%%%%%%%%%%%%%%%%%%%%%%%%%%%%%%%%%%%%%%%
\begin{flushleft}
{\large{\bf 1. Introduction}}
\end{flushleft}
%%%%%%%%%%%%%%%%%%%%%%%%%%%%%%%%%%%%%%%%%%%%%%%%%%%%%%%%%%%%%%%%%%%%%%%%%%%%%
\bigskip
In recent years the study of cohomological gauge theories on manifolds of
special holonomy in various dimensions have attracted a lot of interest
\cite{2,3,4,5}. In the context of string theory and M--theory special
holonomy plays a prominent role especially because the simplest vacua
preserving part of the supersymmetry are compactifications on manifolds of
special holonomy. So far, Calabi--Yau three--folds with $SU(3)$ holonomy
have received the most intensive study in connection with
heterotic string compactifications and also due to the miraculous
mirror symmetry of type II strings on such manifolds. Recently,
the $G_2$-- and $Spin(7)$--holonomy Joyce seven-- and
eight--folds, respectively, have received considerable attention
as well, since they may provide the simplest way to compactify
M--theory to four dimensions and to understand the dynamics of $N = 1$
supersymmetric field theories.

Moreover, it has been shown \cite{3,4} that the ideas underlying
topological theories can be extended to dimensions higher than
four. These cohomological theories, which do not require for a topological
twist, acquire many of the characteristics of a topological theory.
Nevertheless, such theories, which have a rather intriguing structure, are
not fully topological, since they are only invariant under those metric
variations which do not change the {\it reduced} holonomy structure.
At present, the physical status of such theories in $D > 4$ has not been
entirely understood. But, since it is widely believed that the effective
world volume theory of the D--brane is the dimensional reduction of
$N = 1$, $D = 10$ super Yang--Mills theory (SYM) \cite{6}, such cohomological
gauge theories could arise naturally in the study of wrapped Euclidean
D--branes in string theory (see, e.g., \cite{7}).

Examples of such theories in $D = 8$ are the SYM on $Spin(7)$
holonomy Joyce eight--folds \cite{3,4} --- a $N_T = 1$ theory
which is the eight--dimensional analogue of Donaldson--Witten
theory
--- and the SYM on a Calabi--Yau four--folds with $Spin(6) \sim
SU(4)$ holonomy \cite{3} --- a $N_T = 2$ theory which is a
holomorphic analogue of Donaldson--Witten theory. The only other
possible Ricci--flat manifolds admitting covariant constant
(parallel) spinors are hyper--K\"ahler eight--folds with $Spin(5)
\sim Sp(4)$ holonomy which could be interesting as well \cite{8}.
In \cite{1}, making use of a previous work by Ward \cite{9}, Blau
and Thompson gave a group theoretical description of the SYM on
that manifolds. The aim of the present paper is to construct this
theory explicitly. In view of possible generalizations, it might
be instructive to start with the somewhat more involved case of a
quaternionic K\"ahler manifold with $Sp(4) \otimes Sp(2)$ holonomy
and to consider the particular case of a hyper--K\"ahler
eight--fold only in the final analysis.

The paper is organized as follows. In Sect. 2, we first derive, in
accordance with \cite{9}, the generalized $Sp(4) \otimes Sp(2)$
instanton equations. Then, we give the action and the whole set of
transformation rules of the Euclidean $N = 2$, $D = 8$ SYM in flat space
with the original full $SO(8)$ rotation invariance reduced to
$Sp(4) \otimes Sp(2)$. In Sect. 3, we formulate this theory on the
hyper--K\"ahler eight--fold, i.e, for the particular case when the curvature
of the $Sp(2)$ spin connections vanishes. Otherwise, i.e., for the general
case of a quaternionic K\"ahler manifold, which is a Einstein space and not
Ricci--flat, there are no parallel spinors. In the Appendix we describe
in some detail the derivation of the cohomological $N_T = 3$, $D = 8$ SYM.

\bigskip\medskip
%%%%%%%%%%%%%%%%%%%%%%%%%%%%%%%%%%%%%%%%%%%%%%%%%%%%%%%%%%%%%%%%%%%%%%%%%%%%%
\begin{flushleft}
{\large{\bf 2. The $Sp(4) \otimes Sp(2)$--invariant, $N_T = 3$
Euclidean super Yang--Mills theory in eight dimensions}}
\end{flushleft}
%%%%%%%%%%%%%%%%%%%%%%%%%%%%%%%%%%%%%%%%%%%%%%%%%%%%%%%%%%%%%%%%%%%%%%%%%%%%%
\bigskip
Let us first consider the eight--dimensional Euclidean space
$\mathbb R^8$ when the full symmetry group $SO(8)$ is reduced to
its subgroup $Sp(4) \otimes Sp(2)$. Then, the Euclidean
coordinates $x^M$ ($M = 1, \ldots, 8$) can be expressed through
complex coordinates $z_{A a}$, where $A = 1,2,3,4$ and $a = 1,2$
are $Sp(4)$ and $Sp(2)$ indices, respectively, thereby preserving
the Euclidean metric,
\begin{equation*}
d s^2 = d x^M d x_M = \epsilon^{AB} \epsilon^{ab} d z_{A a} d z_{B b},
\qquad
x_M = e_M^{\!~~A a} z_{A a}.
\end{equation*}
Here, $e_M^{\!~~A a}$ defines a (non--singular) map from the
$8$--dimensional Euclidean space to the $4$--dimensional complex
space $\mathbb C^4$,
\begin{equation*}
e_M^{\!~~A a}\, e_{N A a} = \delta_{MN},
 \qquad
 e_M^{\!~~A a}\,e^{M B b} = \epsilon^{AB} \epsilon^{ab}.
\end{equation*}
The indices $A$ and $a$ are raised and lowered as follows, $z^{A
a} = \epsilon^{AB} \epsilon^{ab} z_{B b}$ and $z_{B b} = z^{A a}
\epsilon_{AB} \epsilon_{ab}$, with $\epsilon_{AC} \epsilon^{BC} =
\delta_A^{\!~~B}$ and $\epsilon_{ac} \epsilon^{bc} =
\delta_a^{~b}$, where $\epsilon_{AB}$ and $\epsilon_{ab}$ are the
invariant symplectic tensors of the group $Sp(4)$ and $Sp(2)$,
respectively, $\epsilon_{12} = - \epsilon_{21} = \epsilon_{34} = -
\epsilon_{43} = 1$. Explicitly, for the components of $z_{A a} =
e^M_{\!~~A a} x_M$ we choose
\begin{alignat*}{4}
&z_{11} = x_1 + i x_2,
&\qquad
&z_{21} = - x_3 + i x_4,
&\qquad
&z_{31} = x_5 + i x_6,
&\qquad
&z_{41} = - x_7 + i x_8,
\\
&z_{12} = x_3 + i x_4,
&\qquad
&z_{22} = x_1 - i x_2,
&\qquad
&z_{32} = x_7 + i x_8,
&\qquad
&z_{42} = x_5 - i x_6,
\end{alignat*}
which will be grouped into the complex conjugated coordinates
$z_A \equiv ( z_{11}, z_{12}, z_{31}, z_{32} )$ and
$\bar{z}_A \equiv ( z_{22}, - z_{21}, - z_{42}, z_{41} )$.

Then, on $\mathbb C^4$ an irreducible, $Sp(4) \otimes
Sp(2)$--invariant action of the gauge field $A_A^{\!~~a}$, being
in the adjoint representation of some compact gauge group $G$, is
given by
\begin{equation}
\label{2.1}
S_{\rm YM} = \int d^4z \, d^4\bar{z}\, {\rm tr} \Bigr\{
\hbox{$\frac{1}{4}$} F^{AB}_{\!~~~~ab} F_{AB}^{\!~~~~ab} \Bigr\},
\end{equation}
where $F_{AB}^{\!~~~~ab} = \partial_{[A}^{~~a} A_{B]}^{~~b} + [
A_A^{\!~~a}, A_B^{\!~~b} ]$ is the corresponding field strength.

Besides of (\ref{2.1}), one can construct also a first--stage reducible,
$Sp(4) \otimes Sp(2)$--invariant action, namely
\begin{equation}
\label{2.2}
S_{\rm T} = \int d^4z\, d^4\bar{z}\, {\rm tr} \Bigr\{
\hbox{$\frac{1}{12}$} \Bigr(
\epsilon^{ab} F^{AB}_{\!~~~~ab} \epsilon^{cd} F_{AB cd} -
\epsilon_{AB} F^{AB}_{\!~~~~ab} \epsilon_{CD} F^{CD ab} -
F^{AB}_{\!~~~~ab} F_{AB}^{\!~~~~ba} \Bigr) \Bigr\},
\end{equation}
which can be recast into the form
\begin{equation*}
S_{\rm T} = \int d^4z\, d^4\bar{z}\, {\rm tr} \Bigr\{
\hbox{$\frac{1}{24}$} \epsilon_{ABCD}^{\!~~~~~~~~abcd}
F^{AB}_{\!~~~~ab} F^{CD}_{\!~~~~cd} \Bigr\},
\end{equation*}
where the (fourth rank) tensor $\epsilon_{ABCD}^{\!~~~~~~~~abcd}$
is totally skew--symmetric in the index pairs $(A a)$, $(B b)$,
$(C c)$ and $(D d)$,
\begin{equation}
\label{2.3}
\epsilon_{ABCD}^{\!~~~~~~~~abcd} =
\epsilon_{AB} \epsilon_{CD} \Omega^{cbad} +
\epsilon_{BC} \epsilon_{AD} \Omega^{acbd} +
\epsilon_{CA} \epsilon_{BD} \Omega^{bacd},
\end{equation}
thereby, for later use, we have introduced the following projection operator,
\begin{equation}
\label{2.4}
\Omega_{abcd} = \epsilon_{ac} \epsilon_{bd} + \epsilon_{ad} \epsilon_{bc},
\qquad
\hbox{$\frac{1}{2}$} \Omega_{abef} \Omega^{ef}_{~~cd} = \Omega_{abcd}.
\end{equation}
The tensor (\ref{2.3}) defines a linear map of the space of field
strengths, $F_{AB}^{\!~~~~ab}$, onto itself, namely
$\hbox{$\frac{1}{2}$} \epsilon_{ABCD}^{\!~~~~~~~~abcd}
F^{CD}_{\!~~~~cd} = \lambda F_{AB}^{\!~~~~ab}$. Thus, one can look
for its eigenvalues $\lambda$. In accordance with the $Sp(4)
\otimes Sp(2)$ decomposition of the adjoint representation of
$SO(8)$, $\mathbf{28} \rightarrow \mathbf{(4,2)} \otimes
\mathbf{(4,2)} = \mathbf{(1,3)} \oplus \mathbf{(5,3)} \oplus
\mathbf{(10,1)}$ \cite{9,1}, it is straightforward to prove that
the 3 irreducible subspaces of $\left\{F_{AB}^{\!~~~~ab}\right\}$
are characterized by the eigenvalues $\lambda = -5$, $-1$ and $3$.
Accordingly, under the branching $SO(8) \rightarrow Sp(4) \otimes
Sp(2)$ the field strength decomposes into
\begin{equation}
\label{2.5}
F_{MN} \rightarrow F_{AB}^{\!~~~~ab} = \hbox{$\frac{1}{4}$} \epsilon_{AB}
\epsilon_{CD} F^{CD ab} +
\hbox{$\frac{1}{4}$} \Omega_{ABCD} F^{CD ab} +
\hbox{$\frac{1}{2}$} \epsilon^{ab} \epsilon^{cd} F^{AB}_{\!~~~~cd},
\end{equation}
where, besides (\ref{2.4}), we still have introduced another projection
operator,
\begin{equation}
\label{2.6}
\Omega_{ABCD} = 2 ( \epsilon_{AC} \epsilon_{BD} -
\epsilon_{AD} \epsilon_{BC} ) - \epsilon_{AB} \epsilon_{CD},
\qquad
\hbox{$\frac{1}{4}$} \Omega_{ABEF} \Omega^{EF}_{\!~~~~CD} = \Omega_{ABCD}.
\end{equation}
Thereby, $\epsilon_{CD} F^{CD ab}$ and $\epsilon^{cd}
F^{AB}_{\!~~~~cd}$ in the first and third term on the r.h.s. of
(\ref{2.5}), by virtue of $F_{AB}^{\!~~~~ab} = -
F_{BA}^{\!~~~~ba}$, are symmetric in $(ab)$ and $(AB)$,
respectively, and $\Omega_{ABCD} F^{CD}_{\!~~~~cd}$ in the second
term is symmetric in $(ab)$ and skew--symmetric and trace free in
$(AB)$.

On the other hand, the action (\ref{2.1}) can be rewritten as
\begin{equation}
\label{2.7}
S_{\rm YM} = S_{\rm T} + \int d^4z\, d^4\bar{z}\, {\rm tr} \Bigr\{
\hbox{$\frac{1}{12}$} \Omega_{ABCD} F^{AB}_{\!~~~~ab} F^{CD ab} +
\hbox{$\frac{1}{6}$} \epsilon_{AB} F^{AB}_{\!~~~~ab}
\epsilon_{CD} F^{CD ab} \Bigr\}.
\end{equation}
Hence, if we, formally, just as in case of the $Spin(3)$ instantons
in $D = 4$, impose $S_{\rm YM} = S_{\rm T}$, by virtue of (\ref{2.5}),
it follows that the $Sp(4) \otimes Sp(2)$ instanton equations are
characterized by the eigenvalue $\lambda = 3$,
\begin{equation}
\label{2.8}
\hbox{$\frac{1}{6}$} \epsilon_{ABCD}^{\!~~~~~~~~abcd} F^{CD}_{\!~~~~cd} =
F_{AB}^{\!~~~~ab}
\qquad
\Leftrightarrow
\qquad
\Omega_{ABCD} F^{CD}_{\!~~~~ab} = 0,
\qquad
\epsilon_{AB} F^{AB}_{\!~~~~ab} = 0.
\end{equation}
This result is in accordance with \cite{9}, where the $Sp(4)
\otimes Sp(2)$--invariant integrable equations for the gauge field
$A_A^{\!~~a}$ are discussed, and where the 18 equations
(\ref{2.8}) are obtained as the integrability conditions $\pi_a
\pi_b F_{AB}^{\!~~~~ab} = 0$ of the linear equations $\pi_a (
\partial_A^{\!~~a} + A_A^{\!~~a}) \psi = 0$; here, $\pi_a$ are the
homogeneous coordinates on the complex projective space,
$\mathbb{CP}$.
\bigskip

Now, we like to construct a cohomological theory, whose action localizes
onto the moduli space of the generalized self--duality equations (\ref{2.8}),
in flat space. This theory can be obtained from the Euclidean $N = 2$, $D = 8$
SYM by reducing the group $SO(8)$ to $Sp(4) \otimes Sp(2)$.

The gauge multiplet of the Euclidean $N = 2$, $D = 8$ SYM consists
of the vector field $A_M$, a chiral and anti--chiral Weyl spinor,
$\lambda$ and $\bar{\lambda}$, and the scalar fields $\phi$ and
$\bar{\phi}$. The vector, the chiral spinor and the anti--chiral
spinor representation, which are all eight--dimensional, will be
denoted by $\mathbf{8}_{\bf v}$, $\mathbf{8}_{\bf s}$ and
$\mathbf{8}_{\bf c}$, respectively. Under the branching $SO(8)
\rightarrow Sp(4) \otimes Sp(2)$ these representations decompose
as follows \cite{1},
\begin{alignat}{2}
\label{2.9}
&\mathbf{8}_{\bf v} \rightarrow \mathbf{(4,2)},
&\qquad
&A_M \rightarrow A_A^{\!~~a},
\nonumber
\\
&\mathbf{8}_{\bf s} \rightarrow \mathbf{(5,1)} \oplus \mathbf{(1,3)},
&\qquad
&\lambda \rightarrow \chi_{AB}, \eta^{ab},
\nonumber
\\
&\mathbf{8}_{\bf c} \rightarrow \mathbf{(4,2)},
&\qquad
&\bar{\lambda} \rightarrow \psi_A^{\!~~a},
\end{alignat}
i.e., the vector and the anti--chiral representation remain
irreducible whereas the the spinor representation decomposes into
$\chi_{AB}$ which is skew--symmetric and traceless, $\epsilon^{AB}
\chi_{AB} = 0$, and $\eta^{ab}$ which is symmetric.

After reducing $SO(8)$ to $Sp(4) \otimes Sp(2)$ one ends up with
the following $Sp(4) \otimes Sp(2)$--invariant action with an
extended, $N_T = 3$, {\it on--shell} equivariantly nilpotent
topological supersymmetry (for details, see, Appendix),
\begin{align}
\label{2.10} S^{(N_T = 3)}_{\bigr|Sp(4) \otimes Sp(2) \subset
SO(8)} = \int d^4z\, d^4\bar{z}\, {\rm tr} \Bigr\{&
\hbox{$\frac{1}{4}$} F^{AB}_{\!~~~~ab} F_{AB}^{\!~~~~ab} - 2 D^{A
a} \bar{\phi} D_{A a} \phi \nonumber
\\
& - 2 \chi_{AB} D^{A a} \psi^B_{\!~~a} +
2 \eta^{ab} D^A_{\!~~a} \psi_{A b}  - 2 [ \bar{\phi}, \phi ]^2
\nonumber
\\
& - 2 \bar{\phi} \{ \psi^{A a}, \psi_{A a} \} -
\hbox{$\frac{1}{2}$} \phi \{ \chi^{AB}, \chi_{AB} \} -
\phi \{ \eta^{ab}, \eta_{ab} \} \Bigr\},
\end{align}
where $D_A^{\!~~a} = \partial_A^{\!~~a} + [ A_A^{\!~~a}, ~\cdot~ ]$.
All the fields are in the adjoint representation and take their values
in the Lie algebra $Lie(G)$ of the gauge group $G$.

Furthermore, denoting the 16 (real) supercharges by $Q^{ab}$,
$Q_{AB}$ and $\bar{Q}_A^{\!~~a}$, where $Q^{ab}$ is symmetric,
$Q^{ab} = \frac{1}{2} \Omega^{abcd} Q_{cd}$, and $Q_{AB}$ is
skew--symmetric and traceless, $Q_{AB} = \frac{1}{4} \Omega_{ABCD}
Q^{CD}$, for the {\it on--shell} $Sp(4)$ scalar, tensor and vector
supersymmetry transformations one gets
\begin{align}
\label{2.11}
&Q^{ab} A_A^{\!~~c} = \Omega^{abcd} \psi_{A d},
\nonumber
\\
&Q^{ab} \psi_A^{\!~~c} = \Omega^{abcd} D_{A d} \phi,
\nonumber
\\
&Q^{ab} \phi = 0,
\nonumber
\\
\nonumber
&Q^{ab} \bar{\phi} = \eta^{ab},
\nonumber
\\
&Q^{ab} \eta^{cd} = - \hbox{$\frac{1}{4}$} \Omega^{abe}_{\!~~~~g}
\Omega^{cdfg} \epsilon_{AB} F^{AB}_{\!~~~~ef} +
\Omega^{abcd} [ \phi, \bar{\phi} ],
\nonumber
\\
&Q^{ab} \chi_{AB} = \hbox{$\frac{1}{4}$}
\Omega^{abcd} \Omega_{ABCD} F^{CD}_{\!~~~~cd},
\\
\nonumber\\ %\intertext{}
\label{2.12}
&Q_{AB} A_C^{\!~~a} = \Omega_{ABCD} \psi^{D a},
\nonumber
\\
&Q_{AB} \psi_C^{\!~~a} = - \Omega_{ABCD} D^{D a} \phi,
\nonumber
\\
&Q_{AB} \phi = 0,
\nonumber
\\
\nonumber
&Q_{AB} \bar{\phi} = \chi_{AB},
\nonumber
\\
&Q_{AB} \eta^{ab} = - \hbox{$\frac{1}{4}$} \Omega_{ABCD}
\Omega^{abcd} F^{CD}_{\!~~~~cd},
\nonumber
\\
&Q_{AB} \chi_{CD} = - \hbox{$\frac{1}{4}$} \Omega_{ABEG}
\Omega_{CDF}^{~~~~~~G} \epsilon^{ab} F^{EF}_{\!~~~~ab} +
\Omega_{ABCD} [ \phi, \bar{\phi} ],
\\
\intertext{and}
\label{2.13}
&\bar{Q}_A^{\!~~a} A_B^{\!~~b} = \epsilon_{AB} \eta^{ab} +
\epsilon^{ab} \chi_{AB},
\nonumber
\\
&\bar{Q}_A^{\!~~a} \psi_B^{\!~~b} = F_{AB}^{\!~~~~ab} -
\hbox{$\frac{1}{4}$} \Omega_{ABCD} \Omega^{abcd} F^{CD}_{\!~~~~cd} +
\epsilon_{AB} \epsilon^{ab} [ \phi, \bar{\phi} ],
\nonumber
\\
&\bar{Q}_A^{\!~~a} \phi = - \psi_A^{\!~~a},
\nonumber
\\
&\bar{Q}_A^{\!~~a} \bar{\phi} = 0,
\nonumber
\\
&\bar{Q}_A^{\!~~a} \eta^{cd} = - \Omega^{abcd} D_{A b} \bar{\phi},
\nonumber
\\
&\bar{Q}_A^{\!~~a} \chi_{CD} = \Omega_{ABCD} D^{B a} \bar{\phi},
\end{align}
respectively, where the projection operators $\Omega^{abcd}$ and
$\Omega_{ABCD}$ have been introduced in (\ref{2.4}) and
(\ref{2.6}).

The {\it on--shell} algebraic relations among the supercharges $Q^{ab}$,
$Q_{AB}$ and $\bar{Q}_A^{\!~~a}$ are
\begin{gather*}
\{ Q^{ab}, Q^{cd} \} \doteq - 2 \Omega^{abcd} \delta_G(\phi),
\qquad
\{ Q_{AB}, Q_{CD} \} \doteq 2 \Omega_{ABCD} \delta_G(\phi),
\\
\{ Q^{ab}, Q_{AB} \} \doteq 0,
\\
\{ Q^{ab}, \bar{Q}_A^{\!~~c} \} \doteq - \Omega^{abcd} (
\partial_{A d} + \delta_G(A_{A d}) ), \qquad \{ Q_{AB},
\bar{Q}_C^{\!~~a} \} \doteq \Omega_{ABCD} ( \partial^{D a} +
\delta_G(A^{D a}) ),
\\
\{ \bar{Q}_A^{\!~~a}, \bar{Q}_B^{\!~~b} \} \doteq
2 \epsilon_{AB} \epsilon^{ab} \delta_G(\bar{\phi}),
\end{gather*}
where $\doteq$ indicates that the corresponding equations hold
on--shell, and $\delta_G(\varphi)$ denotes a gauge transformation
depending on the fields $\varphi = ( A_A^{\!~~a}, \phi, \bar{\phi}
)$, being defined by $\delta_G(\varphi) A_A^{\!~~a} = -
D_A^{\!~~a}(A) \varphi$ and $\delta_G(\varphi) X = [ \varphi, X ]$
for all the other fields.

In order to verify that the the action (\ref{2.10}) is invariant
under the transformations (\ref{2.11})--(\ref{2.13}) one has to
use the following basic relations for the projection operators,
\begin{align*}
\Omega_{abeg} \Omega_{cdf}^{~~~~\!g} +
\Omega_{abfg} \Omega_{cde}^{~~~~\!g} &=
\Omega_{abce} \epsilon_{df} + \Omega_{abde} \epsilon_{cf} -
\Omega_{cdae} \epsilon_{bf} - \Omega_{cdbe} \epsilon_{af},
\\
\Omega_{abeg} \Omega_{cdf}^{~~~~\!g} -
\Omega_{abfg} \Omega_{cde}^{~~~~\!g} &= 2 \Omega_{abcd} \epsilon_{ef},
\\
\intertext{and}
\Omega_{ABEG} \Omega_{CDF}^{~~~~~~G} +
\Omega_{ABFG} \Omega_{CDE}^{~~~~~~G} &=
\Omega_{ABCE} \epsilon_{DF} - \Omega_{ABDE} \epsilon_{CF} +
\Omega_{ABCF} \epsilon_{DE} - \Omega_{ABDF} \epsilon_{CE}
\\
&\quad - \Omega_{CDAE} \epsilon_{BF} + \Omega_{CDBE} \epsilon_{AF} -
\Omega_{CDAF} \epsilon_{BE} + \Omega_{CDBF} \epsilon_{AE},
\\
\Omega_{ABEG} \Omega_{CDF}^{~~~~~~G} -
\Omega_{ABFG} \Omega_{CDE}^{~~~~~~G} &= 2 \Omega_{ABCD} \epsilon_{EF},
\end{align*}
thereby we used the following equalities,
\begin{gather*}
\epsilon_{ac} \epsilon_{bd} + \hbox{cyclic}~(a,b,c) = 0,
\\
\epsilon_{AC} \epsilon_{BD} \epsilon_{EF} -
\epsilon_{AC} ( \epsilon_{BE} \epsilon_{DF} - \epsilon_{BF} \epsilon_{DE} ) -
\epsilon_{BD} ( \epsilon_{AE} \epsilon_{CF} - \epsilon_{AF} \epsilon_{CE} ) +
\hbox{cyclic}~(A,B,C) = 0.
\end{gather*}

Finally, in order to obtain a cohomological action we still have
to split off from (\ref{2.10}) the first--stage reducible action
(\ref{2.2}),
\begin{equation*}
S^{(N_T = 3)} = S_{\bigr|Sp(4) \otimes Sp(2) \subset SO(8)}^{(N_T
= 3)} - S_{\rm T},
\end{equation*}
which, by virtue of (\ref{2.7}), yields
\begin{align}
\label{2.14}
S^{(N_T = 3)} = \int d^4z\, d^4\bar{z}\, {\rm tr} \Bigr\{&
\hbox{$\frac{1}{12}$} \Omega_{ABCD} F^{AB}_{\!~~~~ab} F^{CD ab} +
\hbox{$\frac{1}{6}$} \epsilon_{AB} F^{AB}_{\!~~~~ab}
\epsilon_{CD} F^{CD ab}
\nonumber
\\
& - 2 \chi_{AB} D^{A a} \psi^B_{\!~~a} +
2 \eta^{ab} D^A_{\!~~a} \psi_{A b} -
2 \bar{\phi} \{ \psi^{A a}, \psi_{A a} \}
\nonumber
\\
& - \hbox{$\frac{1}{2}$} \phi \{ \chi^{AB}, \chi_{AB} \} -
\phi \{ \eta^{ab}, \eta_{ab} \} - 2 D^{A a} \bar{\phi} D_{A a} \phi -
2 [ \bar{\phi}, \phi ]^2 \Bigr\}.
\end{align}
Then, {\it on--shell}, upon using the equations of motion of $\chi_{AB}$
and $\eta^{ab}$, the action (\ref{2.14}) can be cast into the $Q^{ab}$--exact
form
\begin{equation*}
S^{(N_T = 3)} \doteq Q^{ab} \Psi_{ab},
\end{equation*}
with the gauge fermion
\begin{equation*}
\Psi_{ab} = \Omega_{abcd} \int d^4z\, d^4\bar{z}\, {\rm tr} \Bigr\{
\hbox{$\frac{1}{12}$} \chi_{AB} F^{AB cd} +
\hbox{$\frac{1}{12}$} \eta^c_{\!~~e} \epsilon_{AB} F^{AB de} +
\hbox{$\frac{1}{6}$} \eta^{cd} [ \phi, \bar{\phi} ] -
\hbox{$\frac{1}{3}$} \bar{\phi} D^{A c} \psi_A^{\!~~d} \Bigr\}.
\end{equation*}

\bigskip\medskip
%%%%%%%%%%%%%%%%%%%%%%%%%%%%%%%%%%%%%%%%%%%%%%%%%%%%%%%%%%%%%%%%%%%%%%%%%%%%%
\begin{flushleft}
{\large{\bf 3. Euclidean super Yang--Mills theory on the
hyper--K\"ahler eightfold}}
\end{flushleft}
%%%%%%%%%%%%%%%%%%%%%%%%%%%%%%%%%%%%%%%%%%%%%%%%%%%%%%%%%%%%%%%%%%%%%%%%%%%%%
\bigskip
So far we have assumed that the space is flat, but we shall now formulate
the cohomological theory on a hyper--K\"ahler eight--fold. In view of
possible generalizations it may be convenient to study the more involved
case of a quaternionic K\"ahler manifold. For that purpose, we shall begin
by considering an eight--dimensional Riemannian manifold with
$Sp(4) \otimes Sp(2)$ holonomy,
endowed with hermitean metric
$g_{\mu\nu}$ and $Sp(4)$ and $Sp(2)$ spin connections $\omega_\mu^{\!~~AB}$
and $\omega_\mu^{\!~~ab}$, respectively.
The curved coordinates will be indicated as $x^\mu$ ($\mu = 1, \ldots, 8$)
and the complex coordinates will again be denoted by $z_{A a}$,
\begin{equation*}
d s^2 = g^{\mu\nu} d x_\mu d x_\nu,
\qquad
x_\mu = e_\mu^{\!~~A a} z_{A a},
\end{equation*}
where, locally, $e_\mu^{\!~~A a}$ is the invertible $8$--bein on
the quaternionic K\"ahler manifold,
\begin{equation*}
e_\mu^{\!~~A a}\, e_{\nu A a} = g_{\mu\nu},
 \qquad
 e_\mu^{\!~~A a}\,e^{\mu B b} = \epsilon^{AB} \epsilon^{ab}.
\end{equation*}
In order to break down the structure group $GL(4,\mathbb{C}) \otimes
GL(2,\mathbb{C})$ to $Sp(4) \otimes Sp(2)$, we have to require covariant
constancy of the symplectic tensors $\epsilon^{AB}$ and
$\epsilon^{ab}$,
\begin{align*}
&\nabla_\mu \epsilon^{AB} \equiv \partial_\mu \epsilon^{AB}
 + \omega_{\mu~~C}^{\!~~A}\,\epsilon^{CB}
 + \omega_{\mu~~C}^{\!~~B}\,\epsilon^{AC} = 0,
\\
&\nabla_\mu \epsilon^{ab} \equiv \partial_\mu \epsilon^{ab}
 + \omega_{\mu\!~~c}^{\!~~a}\, \epsilon^{cb}
 + \omega_{\mu\!~~c}^{\!~~b}\, \epsilon^{ac} = 0,
\end{align*}
where $\nabla_\mu$ denotes the metric covariant derivative. In
addition, the integrability conditions $[ \nabla_\mu, \nabla_\nu ]
( \epsilon^{AB}, \epsilon^{ab} ) = 0$ imply that $\epsilon^{AB}$
and $\epsilon^{ab}$ are constant and that the antisymmetric part
of the spin connections can be chosen equal to zero.

Furthermore, in order to ensure covariant constancy of the metric, we impose
\begin{equation}
\label{3.1}
\nabla_\mu e_\nu^{\!~~A a} \equiv \partial_\mu e_\nu^{\!~~A a} -
\Gamma_{\mu\nu}^{~~~\lambda} e_\lambda^{\!~~A a}
 +
\omega_{\mu~~B}^{\!~~A}\, e_\nu^{\!~~B a}
 +
\omega_{\mu\!~~b}^{\!~~a}\, e_\nu^{\!~~A b} = 0,
\end{equation}
where $\Gamma_{\mu\nu}^{~~~\lambda}$ is the affine connection.

The crucial ingredient of a quaternionic K\"ahler manifold is a
triplet of complex structures,
\begin{equation}
\label{3.2} (J^\alpha)_\mu^{\!~~\nu} = - i e_\mu^{\!~~A a}
(\sigma^\alpha)_a^{\!~~b} \, e^\nu_{\!~~A b}, \qquad \alpha =
1,2,3,
\end{equation}
where $(\sigma^\alpha)_a^{\!~~b}$ are the $Sp(2)$ generators,
$(\sigma^\alpha)_a^{\!~~c} (\sigma^\beta)_c^{\!~~b} =
i \epsilon^{\alpha\beta\gamma} (\sigma_\gamma)_a^{\!~~b} +
\delta^{\alpha\beta} \delta_a^{\!~~b}$, which obey the algebra of the
quaternions,
\begin{equation*}
(J^\alpha)_\mu^{\!~~\rho} (J^\beta)_\rho^{\!~~\nu} =
\epsilon^{\alpha\beta\gamma} (J_\gamma)_\mu^{\!~~\nu} -
\delta_\mu^{\!~~\nu} \delta^{\alpha\beta}.
\end{equation*}
Since the metric $g_{\mu\nu}$ is preserved and hermitean, it holds
\begin{equation*}
(J^\alpha)_\mu^{\!~~\rho} g_{\rho\nu} +
(J^\alpha)_\nu^{\!~~\rho} g_{\rho\mu} = 0.
\end{equation*}

Now, for any choice of $(J_\alpha)_\mu^{\!~~\nu}$, we can
associate a triplet of K\"ahler two--forms $\rho_\alpha$ via
\begin{equation*}
\rho_\alpha \equiv \hbox{$\frac{1}{2}$} i
(J_\alpha)_\mu^{\!~~\rho} g_{\rho\nu} dx^\mu \wedge dx^\nu =
\hbox{$\frac{1}{2}$} \epsilon_{AB} (\sigma_\alpha)^{ab}
dz^A_{\!~~a} \wedge dz^B_{\!~~b}.
\end{equation*}
Then, by making use of the completeness relation,
$(\sigma_\alpha)_a^{\!~~c} (\sigma^\alpha)_b^{\!~~d} =
\epsilon_{ab} \epsilon^{cd} - \delta_a^{\!~~d} \delta_b^{\!~~c}$,
we can define the four--form
\begin{equation*}
\Omega \equiv \hbox{$\frac{1}{2}$} \rho_\alpha \wedge \rho^\alpha =
\hbox{$\frac{1}{24}$} \epsilon_{ABCD}^{\!~~~~~~~~abcd}
dz^A_{\!~~a} \wedge dz^B_{\!~~b} \wedge dz^C_{\!~~c} \wedge dz^D_{\!~~d},
\end{equation*}
which reveals the geometrical origin of the fourth rank tensor
$\epsilon_{ABCD}^{\!~~~~~~~~abcd}$ introduced in (\ref{2.3}).

Furthermore, on a quaternionic K\"ahler manifold the complex
structures (\ref{3.2}) are required to be covariant constant as
well,
\begin{equation}
\label{3.3}
\nabla_\mu (J^\alpha)_\nu^{\!~~\lambda} \equiv
\partial_\mu (J^\alpha)_\nu^{\!~~\lambda} -
\Gamma_{\mu\nu}^{~~~\rho} (J^\alpha)_\rho^{\!~~\lambda} +
\Gamma_{\mu\rho}^{~~~\lambda} (J^\alpha)_\nu^{\!~~\rho} +
2 \epsilon^{\alpha\beta\gamma} \omega_{\mu \beta}
(J_\gamma)_\nu^{\!~~\lambda} = 0,
\end{equation}
where in place of $\omega_\mu^{~ab}$ we have used the triplet
representation $\omega_\mu^{\!~~\alpha} = \frac{1}{2} i
(\sigma^\alpha)_{ab} \omega_\mu^{\!~~ab}$. Since a quaternionic
K\"ahler manifold --- in contrast to a hyper--K\"ahler eight--fold
--- does not have a vanishing Nijenhuis tensor, the affine
connection and the $Sp(2)$ spin connection can not be uniquely
defined without additional requirements. \footnote{The covariant
constancy condition (\ref{3.3}) is left invariant when one
performs simultaneously the replacements
$\Gamma_{\mu\nu}^{~~~\lambda} \rightarrow
\Gamma_{\mu\nu}^{~~~\lambda} + ( \delta_{(\mu}^{~~\rho}
\delta_{\nu)}^{~~\lambda} - (J^\alpha)_{(\mu}^{~~\rho}
(J_\alpha)_{\nu)}^{~~\lambda} ) \xi_\rho$ and
$\omega_\mu^{\!~~\alpha} \rightarrow \omega_\mu^{\!~~\alpha} +
(J^\alpha)_\mu^{\!~~\rho} \xi_\rho$, where $\xi_\rho$ is an
arbitrary vector (see, Appendix B of \cite{10}). Notice, that in
our convention (anti)symmetrization are defined by $( X, Y ) = X Y
+ X Y$ resp. $[ X, Y ] = X Y - X Y$, whereas in \cite{10} the
corresponding definitions include an additional factor $1/2$.}

In order to define in (\ref{3.3}) the $Sp(2)$ spin connection
$\omega_\mu^{\!~~\alpha}$ we adopt the two requirements proposed in \cite{10},
\begin{equation}
\label{3.4}
(J^\alpha)_\mu^{\!~~\nu} \omega_{\nu \alpha} = 0,
\qquad
N_{\mu\nu}^{~~~\lambda} = - \hbox{$\frac{1}{2}$}
(J^\alpha)_{[\mu}^{\!~~~\lambda} \omega_{\nu] \alpha},
\end{equation}
where $N_{\mu\nu}^{~~~\lambda}$ is the Nijenhuis tensor (in the
normalization of \cite{10})
\begin{equation*}
N_{\mu\nu}^{~~~\lambda} \equiv
\hbox{$\frac{1}{12}$} (J^\alpha)_\mu^{\!~~\rho}
\partial_{[\rho} (J_\alpha)_{\nu]}^{~~\lambda} -
\hbox{$\frac{1}{12}$} (J^\alpha)_\nu^{\!~~\rho}
\partial_{[\rho} (J_\alpha)_{\mu]}^{~~\lambda} = - N_{\nu\mu}^{~~~\lambda}.
\end{equation*}
The second condition in (\ref{3.4}), which ensures that
$\Gamma_{\mu\nu}^{~~~\lambda}$ is actually the (torsionless)
Levi--Civita connection, can be easily solved for $\omega_\mu^{\!~~\alpha}$.
One finds
\begin{equation*}
\omega_\mu^{\!~~\alpha} = - \hbox{$\frac{1}{3}$} N_{\mu\nu}^{~~~\lambda}
(J^\alpha)_\lambda^{\!~~\nu}.
\end{equation*}
Then, one can show that the Levi--Civita connection in (\ref{3.3}) is
equal to the Oproiu connection \cite{11},
\begin{equation}
\label{3.5}
\Gamma_{\mu\nu}^{~~~\lambda} = - \hbox{$\frac{1}{6}$}
\partial_{(\mu} (J^\alpha)_{\nu)}^{~~\rho} (J_\alpha)_\rho^{\!~~\lambda} -
\hbox{$\frac{1}{12}$} \epsilon^{\alpha\beta\gamma} (J_\beta)_{(\mu}^{~~\rho}
\partial_{|\rho|} (J_\gamma)_{\nu)}^{~~\sigma}
(J_\alpha)_\sigma^{\!~~\lambda} - \hbox{$\frac{1}{2}$}
(J^\alpha)_{(\mu}^{\!~~~\lambda} \omega_{\nu) \alpha}.
\end{equation}
It differs from the Obata connection \cite{12} --- the
Levi--Civita connection in the case of a hyper--K\"ahler
eight--fold --- by the $\omega_\mu^{\!~~\alpha}$--dependent term.
With that choice for $\omega_\mu^{\!~~ab} = i (\sigma_\alpha)^{ab}
\omega_\mu^{\!~~\alpha}$ and $\Gamma_{\mu\nu}^{~~~\lambda}$ the
$Sp(4)$ spin connection $\omega_\mu^{\!~~AB}$ can be immediately
obtained from the condition (\ref{3.1}).

>From the integrability condition $[ \nabla_\mu, \nabla_\nu ]
e_\rho^{\!~~A a} = 0$ of (\ref{3.1}) it follows that the
Riemannian curvature $R_{\mu\nu\rho}^{\!~~~~~\sigma} =
\partial_{[\mu} \Gamma_{\nu]\rho}^{\!~~~~\sigma} +
\Gamma_{\lambda[\mu}^{\!~~~~\sigma} \Gamma_{\nu]\rho}^{\!~~~~\lambda}$
decomposes into
\begin{equation*}
R_{\mu\nu\rho}^{\!~~~~~\sigma} = - (J_\alpha)_\rho^{\!~~\sigma}
R_{\mu\nu}^{~~~\alpha} - e^\sigma_{\!~~A a} e_{\rho B}^{\!~~~~a}
R_{\mu\nu}^{~~~AB},
\end{equation*}
where $R_{\mu\nu}^{~~~\alpha} = \frac{1}{2} i (\sigma^\alpha)_{ab}
R_{\mu\nu}^{~~~ab}$ and $R_{\mu\nu}^{~~~AB}$ are the $Sp(2)$ and $Sp(4)$
curvatures, respectively,
\begin{align*}
&R_{\mu\nu}^{~~~\alpha} = \partial_{[\mu} \omega_{\nu]}^{~~\alpha}
+ 2 \epsilon^{\alpha\beta\gamma} \omega_{\mu \beta} \omega_{\nu
\gamma},
\\
&R_{\mu\nu}^{~~~AB} = \partial_{[\mu} \omega_{\nu]}^{~~AB} +
\omega_{[\mu\!~~C}^{\!~~A} \omega_{\nu]}^{~~BC}.
\end{align*}
Furthermore, from the integrability condition $[ \nabla_\mu, \nabla_\nu ]
(J^\alpha)_\rho^{\!~~\sigma} = 0$ of (\ref{3.3}), namely
\begin{equation}
\label{3.6}
R_{\mu\nu\rho}^{\!~~~~~\sigma} (J^\alpha)_\lambda^{\!~~\rho} -
R_{\mu\nu\lambda}^{\!~~~~~\rho} (J^\alpha)_\rho^{\!~~\sigma} +
2 \epsilon^{\alpha\beta\gamma} R_{\mu\nu \beta}
(J_\gamma)_\lambda^{\!~~\sigma} = 0,
\end{equation}
together with the relation $\frac{1}{8}
R_{\mu\nu\rho}^{\!~~~~~\sigma} (J^\alpha)_\sigma^{\!~~\rho} =
R_{\mu\nu}^{~~~\alpha}$, it can be shown that the Ricci tensor
$R_{\mu\nu} = R_{\lambda\mu\nu}^{\!~~~~~\lambda}$ is proportional
to the curvature scalar, $R_{\mu\nu} = \hbox{$\frac{1}{8}$}
g_{\mu\nu} R$, and that the $Sp(2)$ curvature is proportional to
the complex structures, $R_{\mu\nu}^{~~~\alpha} =
\hbox{$\frac{1}{64}$} R (J^\alpha)_{\mu\nu}$ \cite{10}, i.e., the
quaternionic K\"ahler manifold is Einstein in contrast to the
hyper--K\"ahler eight--fold which is Ricci--flat.

After having specified the data of a Riemannian manifold with
$Sp(4) \otimes Sp(2)$ holonomy let us now discuss whether or not
the cohomological $N_T = 3$, $D = 8$ SYM can be coupled to such a
background. In the case of a hyper--K\"ahler eight--fold, i.e.,
for a Ricci--flat manifold with vanishing $Sp(2)$ curvature,
$R_{\mu\nu} = - \hbox{$\frac{8}{3}$} (J_\alpha)_\mu^{\!~~\rho}
R_{\rho\nu}^{~~~\alpha} = 0$, there is no problem. As usual, in
order to put a flat space gauge invariant theory on a curved space one
has to covariantize the action\break (\ref{2.14}) via
$\delta_{MN} \rightarrow g_{\mu\nu}$, $d^4z\, d^4\bar{z}
\rightarrow d^4z\, d^4\bar{z}\, \sqrt{g}$ and $D_{A a} \rightarrow
D_{A a}^{\rm cov} = e_{A a}^{~~~\mu} \nabla_\mu + [ A_{A a},~\cdot~ ]$,
where $D_{A a}^{\rm cov}$ denotes the gauge and metric covariant derivative.
Furthermore, it that case there is a triplet of parallel spinors
$\zeta_\alpha$ obeying
\begin{equation}
\label{3.8}
\nabla_\mu \zeta_\alpha = 0,
\qquad
\nabla_\mu \equiv \partial_\mu + \hbox{$\frac{1}{4}$}
\omega_\mu^{\!~~ABab} \gamma_{ABab}.
\end{equation}
Thereby, $\gamma_{ABab} = \hbox{$\frac{1}{2}$} [ \gamma_{Aa}, \gamma_{Bb} ]$
are the generators of the holonomy group (with $\gamma_{Aa}$ being
the $Sp(4) \otimes Sp(2)$ Dirac matrices) and
\begin{equation}
\label{3.9}
\omega_\mu^{\!~~ABab} = - e^{\nu Aa} ( \partial_\mu
e_\nu^{\!~~Bb} - \Gamma_{\mu\nu}^{~~~\lambda} e_\lambda^{\!~~Bb} )
\end{equation}
is the affine spin connection (with $\Gamma_{\mu\nu}^{~~~\lambda}$
being the Obata connection). The spinors $\zeta_\alpha$ can be
identified with the parameters $\zeta^{ab}$, via $\zeta^{ab} = i
(\sigma^\alpha)^{ab} \zeta_\alpha$, of the scalar supersymmetries
(\ref{2.11}), which are already in a covariantized form.

Obviously, in the case of a quaternionic K\"ahler manifold, i.e.,
if $\Gamma_{\mu\nu}^{~~~\lambda}$ in (\ref{3.9}) agrees with the
Oproiu connection (\ref{3.5}), the integrability condition $[
\nabla_\mu, \nabla_\nu ] \zeta_\alpha = 0$ allows only for the
trivial solution $\zeta_\alpha = 0$, since this manifold is
Einstein, $R \neq 0$. But, in the particular case when the
curvature scalar $R = \frac{7}{2} \lambda_0^2$ is constant (the
pre-factor is dimension dependent), the integrability condition
does not forbid the existence of a triplet of (conformal) Killing
spinors $\zeta_\alpha$ satisfying
\begin{equation}
\label{3.10}
\nabla_\mu \zeta_\alpha = \hbox{$\frac{1}{8}$} \lambda_0
\gamma_\mu \zeta_\alpha,
\qquad
\gamma_\mu = e_\mu^{\!~~Aa} \gamma_{Aa}.
\end{equation}
Indeed, acting on (\ref{3.10}) with $\nabla_\nu$ and antisymmetrising,
by virtue of (\ref{3.9}), one obtains
\begin{equation}
\label{3.11}
e^{\rho Aa} e_\sigma^{\!~~Bb} R_{\mu\nu\rho}^{\!~~~~~\sigma}
\gamma_{ABab} \zeta_\alpha = - \hbox{$\frac{1}{8}$} \lambda_0^2
\gamma_{\mu\nu} \zeta_\alpha,
\qquad
\gamma_{\mu\nu} = e_\mu^{\!~~Aa} e_\nu^{\!~~Bb} \gamma_{ABab},
\end{equation}
where $R_{\mu\nu\rho}^{\!~~~~~\sigma}$ is the Riemannian curvature
tensor being introduced above. Since
$\Gamma_{\mu\nu}^{~~~\lambda}$ is torsionless and since
$g_{\mu\nu}$ is preserved, this curvature obeys the cyclicity
property $R_{\mu\nu\rho}^{\!~~~~~\sigma} +
R_{\nu\rho\mu}^{\!~~~~~\sigma} + R_{\rho\mu\nu}^{\!~~~~~\sigma} =
0$ and the exchange property $R_{\mu\nu\rho\sigma} =
R_{\rho\sigma\mu\nu}$. Then, multiplying (\ref{3.11}) by
$\gamma^\nu$ and taking into account these properties one arrives
at
\begin{equation}
\label{3.12}
R_{\rho\mu\nu}^{\!~~~~~\rho} \gamma^\nu \zeta_\alpha =
R_{\mu\nu} \gamma^\nu \zeta_\alpha =
\hbox{$\frac{7}{16}$} \lambda_0^2 \gamma_\mu \zeta_\alpha,
\end{equation}
where $R_{\mu\nu} = \hbox{$\frac{1}{8}$} g_{\mu\nu} R$. This
condition admits a non--trivial solution for $R = \frac{7}{2}
\lambda_0^2$. Obviously, $\lambda_0 \rightarrow 0$ brings us back to
(\ref{3.8}).

Hence, it is interesting to enquire whether the existence of
parallel spinors on a hyper--K\"ahler eight--fold, which is
irreducible and Ricci--flat, implies self--duality of the spin
connection (and conversely), whereas the existence of Killing
spinors on a quaternionic K\"ahler manifold, which is a Einstein
space, is equivalent to certain generalized self--duality
conditions for the spin connections. With another words, it is
possible --- in the case of a quaternionic K\"ahler manifold with
constant $R = \frac{7}{2} \lambda_0^2$ --- to add certain
$R$--dependent terms to the covariantized action (\ref{2.14}) in
such a way that the Killing spinors $\zeta_\alpha$ may be
identified with the parameters $\zeta^{ab}$, via $\zeta^{ab} = i
(\sigma^\alpha)^{ab} \zeta_\alpha$, of the supersymmetries
(\ref{2.11}). Unfortunately, we have not so far been successful in
finding such $R$--dependent terms. Nevertheless, we believe that
this problem is worthy of further investigations.
\bigskip\medskip
\pagebreak

%%%%%%%%%%%%%%%%%%%%%%%%%%%%%%%%%%%%%%%%%%%%%%%%%%%%%%%%%%%%%%%%%%%%%%%%%%%%%
\begin{flushleft}
{\large{\bf Appendix}}
\end{flushleft}
%%%%%%%%%%%%%%%%%%%%%%%%%%%%%%%%%%%%%%%%%%%%%%%%%%%%%%%%%%%%%%%%%%%%%%%%%%%%%
\bigskip
In order to derive from the Euclidean $N = 2$, $D = 8$ SYM a
cohomological theory where the $SO(8)$ invariance is broken down
to $Sp(4) \otimes Sp(2)$ we proceed as follows: First we choose
the standard embedding of $Sp(4) \otimes Sp(2)$ into $SO(8)$,
which is defined as the stability subgroup $Sp(4) \otimes Sp(2)
\subset SO(8)$ of the vector representation according to which we
have the decomposition
\begin{alignat*}{2}
&\mathbf{8}_{\bf v} \rightarrow \mathbf{(5,1)} \oplus \mathbf{(1,3)},
&\qquad
&A_M \rightarrow G_i, V_\alpha,
\\
&\mathbf{8}_{\bf s} \rightarrow \mathbf{(4,2)},
&\qquad
&\lambda \rightarrow \lambda_{A a},
\\
&\mathbf{8}_{\bf c} \rightarrow \mathbf{(4,2)},
&\qquad
&\bar{\lambda} \rightarrow \bar{\lambda}_{A a}.
\end{alignat*}
Usually, one discards this representation since it does not allow
for the existence of a $Sp(4) \otimes Sp(2)$--invariant, totally
skew--symmetric (fourth rank) tensor. But, in the present case,
after having established the $N = 2$, $D = 8$ SYM in that
representation one can simply deduce the structure of the
cohomological $N_T = 3$, $D = 8$ SYM according to the
decomposition (\ref{2.9}).

To begin with, let us first consider the full $SO(8)$--invariant
action of the Euclidean $N = 2$, $D = 8$ SYM, which is obtained
from the Minkowskian $N = 1$, $D = 10$ SYM \cite{13} by ordinary
dimensional reduction and performing a Wick rotation into the
Euclidean space. It is built up from an anti--hermitean vector
field $A_M$, 16--component chiral and anti--chiral Weyl spinors,
$\lambda$ and $\bar{\lambda}$, respectively, and the scalar fields
$\phi$ and $\bar{\phi}$. All the fields take their values in the
Lie algebra $Lie(G)$ of some compact gauge group $G$. As a result,
for the dimensionally reduced Euclidean action one obtains
\begin{align}
S^{(N = 2)} = \int_E d^8x\, {\rm tr} \Bigr\{&
\hbox{$\frac{1}{4}$} F^{MN} F_{MN} +
2 \bar{\lambda} \Gamma^M D_M \lambda - 2 D^M \bar{\phi} D_M \phi
\nonumber
\\
& + 2 \lambda^T C_8^{-1} [ \phi, \lambda ] -
2 \bar{\lambda} C_8 [ \bar{\phi}, \bar{\lambda}^T ] -
2 [ \bar{\phi}, \phi ]^2 \Bigr\},
\tag{A.1}
\end{align}
where $F_{MN} = \partial_{[M} A_{N]} + [ A_M, A_N ]$ and $D_M =
\partial_M + [ A_M, ~\cdot~ ]$. $C_8$ is the charge conjugation
matrix, $C_8^{-1} \Gamma_M C_8 = - \Gamma_M^T$, which can be
chosen to be symmetric. The Dirac matrices $\Gamma_M$ for the
$SO(8)$ spinor representation will be specified below,
$\frac{1}{2} \{ \Gamma_M, \Gamma_N \} = \delta_{MN} I_{16}$.

The action (A.1) is invariant under the following supersymmetry transformations
\begin{align}
&\delta A_M = \bar{\zeta} \Gamma_M \lambda - \bar{\lambda} \Gamma_M \zeta,
\nonumber
\\
&\delta \phi = \bar{\zeta} C_8 \bar{\lambda}^T,
\nonumber
\\
&\delta \lambda = - \hbox{$\frac{1}{4}$} \Gamma^{MN} \zeta F_{MN} +
\Gamma^M C_8 \bar{\zeta}^T D_M \bar{\phi} - \zeta [ \phi, \bar{\phi} ],
\nonumber
\\
&\delta \bar{\phi} = \lambda^T C_8^{-1} \zeta,
\nonumber
\\
&\delta \bar{\lambda} = \hbox{$\frac{1}{4}$} \bar{\zeta} \Gamma^{MN} F_{MN} -
\zeta^T C_8^{-1} \Gamma^M D_M \phi - \bar{\zeta} [ \bar{\phi}, \phi ],
\tag{A.2}
\end{align}
with $\zeta$ and $\bar{\zeta}$ being constant chiral and anti--chiral Weyl
spinors, respectively ($\Gamma_9 \zeta = - \zeta$ with
$\Gamma_9 = \Gamma_1 \cdots \Gamma_8$), and where
$\Gamma_{MN} = \frac{1}{2} [ \Gamma_M, \Gamma_N ]$ are the $SO(8)$ generators.

In order to reduce in the action (A.1) the $SO(8)$ invariance to
$Sp(4) \otimes Sp(2)$, we replace the $SO(8)$ matrices by the
standard embedding $\Gamma_M = ( \Gamma_i, \Gamma_{5 + \alpha} )$
($i = 1, \ldots, 5$, $\alpha = 1,2,3$) of $Sp(4) \otimes Sp(2)$
into $SO(8)$. In this representation $\Gamma_i, \Gamma_{5 + \alpha}$ and
the chirality matrix $\Gamma_9$ are given by
\begin{align}
&\Gamma_i =
\begin{pmatrix} 0 & (\gamma_i)_A^{\!~~B} \delta_a^{\!~~b} \\
(\gamma_i)_A^{\!~~B} \delta_a^{\!~~b} & 0 \end{pmatrix},
\nonumber
\\
&\Gamma_{5 + \alpha} =
\begin{pmatrix} 0 & - i \delta_A^{\!~~B} (\sigma_\alpha)_a^{\!~~b} \\
i \delta_A^{\!~~B} (\sigma_\alpha)_a^{\!~~b} & 0 \end{pmatrix},
\nonumber
\\
&\Gamma_9 = \Gamma_1 \ldots \Gamma_8 =
\begin{pmatrix} - \delta_A^{\!~~B} \delta_a^{\!~~b} & 0 \\
0 & \delta_A^{\!~~B} \delta_a^{\!~~b} \end{pmatrix},
\tag{A.3}
\end{align}
$(\gamma_i)_A^{\!~~B}$ ($A = 1,2,3,4$) and
$(\sigma_\alpha)_a^{\!~~b}$ ($a = 1,2$) being the $Sp(4)$ and
$Sp(2)$ generators, respectively.

The $Sp(2)$ matrices obey
\begin{equation}
(\sigma_\alpha)_a^{\!~~c} (\sigma_\beta)_c^{\!~~b} =
i \epsilon_{\alpha\beta\gamma} (\sigma^\gamma)_a^{\!~~b} +
\delta^{\alpha\beta} \delta_a^{\!~~b},
\tag{A.4}
\end{equation}
and are symmetric $(\sigma_\alpha)_{ab} = (\sigma_\alpha)_{ba}$.
For the $Sp(4)$ matrices we take the particular representation
\begin{equation*}
(\gamma_\alpha)_A^{\!~~B} = \begin{pmatrix}
0 & - i (\sigma_\alpha)_a^{\!~~b} \\
i (\sigma_\alpha)_a^{\!~~b} & 0 \end{pmatrix},
\qquad
(\gamma_4)_A^{\!~~B} = \begin{pmatrix}
0 & \delta_a^{\!~~b} \\ \delta_a^{\!~~b} & 0 \end{pmatrix},
\qquad
(\gamma_5)_A^{\!~~B} = \begin{pmatrix}
\delta_a^{\!~~b} & 0 \\ 0 & - \delta_a^{\!~~b} \end{pmatrix},
\end{equation*}
where $\alpha$ runs over $1,2,3$ (recalling that $Sp(4) \sim Spin(5)$ is the
covering group of $SO(5)$). They satisfy the relations
\begin{align}
&(\gamma_i)_A^{\!~~C} (\gamma_j)_C^{\!~~B} =
(\gamma_{ij})_A^{\!~~B} + \delta_{ij} \delta_A^{\!~~B},
\nonumber
\\
&(\gamma_i)_A^{\!~~C} (\gamma_{mn})_C^{\!~~B} =
\delta_{i[m} (\gamma_{n]})_A^{\!~~B} -
\hbox{$\frac{1}{2}$} \epsilon_{ijkmn} (\gamma^{jk})_A^{\!~~B},
\nonumber
\\
&(\gamma_{ij})_A^{\!~~C} (\gamma_{mn})_C^{\!~~B} =
\delta_{[i[m} (\gamma_{n]j]})_A^{\!~~B} +
\epsilon_{ijkmn} (\gamma^k)_A^{\!~~B} -
\delta_{i[m} \delta_{n]j} \delta_A^{\!~~B},
\tag{A.5}
\end{align}
where $\epsilon_{ijkmn}$ is a totally antisymmetric unit tensor. Here,
the 5 matrices $(\gamma_i)_{AB}$ are skew--symmetric and traceless,
$(\gamma_i)_{AB} = - (\gamma_i)_{BA}$ and $\epsilon^{AB} (\gamma_i)_{AB} = 0$,
whereas the 10 generators $(\gamma_{ij})_{AB}$ of the $Sp(4)$ rotations are
symmetric, $(\gamma_{ij})_{AB} = (\gamma_{ij})_{BA}$.

Then, for the 64 (antisymmetric) generators
$\Gamma_{MN} = ( \Gamma_{ij}, \Gamma_{i,5 + \alpha},
\Gamma_{5 + \alpha,5 + \beta} )$ from (A.3) one obtains
\begin{align}
&\Gamma_{ij} =
\begin{pmatrix} (\gamma_{ij})_A^{\!~~B} \delta_a^{\!~~b} & 0 \\
0 & (\gamma_{ij})_A^{\!~~B} \delta_a^{\!~~b} \end{pmatrix},
\nonumber
\\
&\Gamma_{i,5 + \alpha} =
\begin{pmatrix} i (\gamma_i)_A^{\!~~B} (\sigma_\alpha)_a^{\!~~b} & 0 \\
0 & - i (\gamma_i)_A^{\!~~B} (\sigma_\alpha)_a^{\!~~b} \end{pmatrix},
\nonumber
\\
&\Gamma_{5 + \alpha,5 + \beta} =
\begin{pmatrix} i \delta_A^{\!~~B} \epsilon_{\alpha\beta\gamma}
(\sigma^\gamma)_a^{\!~~b} & 0 \\
0 & i \delta_A^{\!~~B} \epsilon_{\alpha\beta\gamma}
(\sigma^\gamma)_a^{\!~~b} \end{pmatrix}.
\tag{A.6}
\end{align}
A particular and crucial feature of the action (A.1) is that the charge
conjugation matrix $C_8$ can be chosen to be symmetric,
\begin{equation*}
C_8 = \begin{pmatrix} \epsilon_{ab} \epsilon_{AB} & 0 \\
0 & - \epsilon_{ab} \epsilon_{AB} \end{pmatrix},
\qquad
C_8^{-1} \Gamma_M C_8 = - \Gamma_M^T.
\end{equation*}

After having specified the representation of the matrices $\Gamma_M$
let us now determine the action (A.1) and the transformation rules (A.2)
of the $N = 2$, $D = 8$ SYM with the $SO(8)$ rotation invariance reduced
to $Sp(4) \otimes Sp(2)$. To begin with, we write the Weyl spinors
$\lambda$ and $\bar{\lambda}$ as
\begin{equation}
\lambda = - \Gamma_9 \lambda = \begin{pmatrix}
\lambda_{Aa} \\ 0 \end{pmatrix},
\qquad
\bar{\lambda} = \bar{\lambda} \Gamma_9 = ( 0, \bar{\lambda}^{Aa} ),
\tag{A.7}
\end{equation}
which, just as in the Minkowski space, are related by hermitean
conjugation, $\bar{\lambda} = \lambda^\dagger E_8$, with
\begin{equation*}
E_8 = \begin{pmatrix} 0 & \epsilon^{ab} \epsilon^{AB} \\
\epsilon^{ab} \epsilon^{AB} & 0 \end{pmatrix},
\qquad
E_8 \Gamma_M E_8^{-1} = \Gamma_M^\dagger,
\end{equation*}
where we have renamed $\lambda_{Aa}^\dagger = \bar{\lambda}_{Aa}$.

Then, substituting into (A.1) for $\Gamma_M$ the matrices (A.3)
and splitting the gauge field into $A_M = (G_i, V_\alpha)$, one
ends up with the following $Sp(4) \otimes Sp(2)$--invariant action
with an underlying $N = 2$ supersymmetry,
\begin{align}
S^{(N = 2)}_{\bigr|Sp(4) \otimes Sp(2) \subset SO(8)} = \int
d^4z\, d^4\bar{z}\, {\rm tr} \Bigr\{& \hbox{$\frac{1}{4}$} F^{ij}
F_{ij} + \hbox{$\frac{1}{2}$} F^{i \alpha} F_{i \alpha} +
\hbox{$\frac{1}{4}$} F^{\alpha\beta} F_{\alpha\beta} \nonumber
\\
& + 2 \bar{\lambda}^{Aa} (\gamma^i)_A^{\!~~B} D_i \lambda_{Ba} -
2 D^i \bar{\phi} D_i \phi
\nonumber
\\
& + 2 i \bar{\lambda}^{Aa} (\sigma^\alpha)_a^{\!~~b} D_\alpha \lambda_{Ab} -
2 D^\alpha \bar{\phi} D_\alpha \phi
\nonumber
\phantom{\frac{1}{2}}
\\
& - 2 \phi \{ \lambda_{Aa}, \lambda^{Aa} \} -
2 \bar{\phi} \{ \bar{\lambda}^{Aa}, \bar{\lambda}_{Aa} \} -
2 [ \bar{\phi}, \phi ]^2 \Bigr\}.
\tag{A.8}
\end{align}
For the sake of simplicity, the various field strength tensors and
covariant derivatives are denoted by $F_{ij} = \partial_{[i}
G_{j]} + [ G_i, G_j ]$, $F_{i \alpha} = \partial_i V_\alpha -
\partial_\alpha G_i + [ G_i, V_\alpha ]$, $F_{\alpha\beta} =
\partial_{[\alpha} V_{\beta]} + [ V_\alpha, V_\beta ]$ and\break $D_i =
\partial_i + [ G_i, ~\cdot~ ]$, $D_\alpha = \partial_\alpha + [
V_\alpha, ~\cdot~ ]$, respectively.

Denoting the 16 (real) supercharges with $\bar{Q}_{Aa}$ and $Q^{Aa}$, and
decomposing the supersymmetry transformations according to
$\delta = \bar{\zeta}^{Aa} \bar{Q}_{Aa} - \zeta_{Aa} Q^{Aa}$, from (A.2)
one obtains
\begin{align}
&Q^{Aa} G_i = (\gamma_i)^A_{\!~~B} \bar{\lambda}^{Ba},
\nonumber
\\
&Q^{Aa} V_\alpha = - i (\sigma_\alpha)^a_{\!~~b} \bar{\lambda}^{Ab},
\nonumber
\\
&Q^{Aa} \phi = 0,
\nonumber
\\
&Q^{Aa} \lambda_{Bb} = \hbox{$\frac{1}{4}$} (\gamma_{ij})^A_{\!~~B}
\delta^a_{\!~~b} F^{ij} -
\hbox{$\frac{1}{2}$} i (\gamma_i)^A_{\!~~B} (\sigma_\alpha)^a_{\!~~b}
F^{i \alpha} + \hbox{$\frac{1}{4}$} i \delta^A_{\!~~B}
\epsilon_{\alpha\beta\gamma} (\sigma^\gamma)^a_{\!~~b} F^{\alpha\beta} +
\delta^A_{\!~~B} \delta^a_{\!~~b} [ \phi, \bar{\phi} ],
\nonumber
\\
&Q^{Aa} \bar{\phi} = \lambda^{Aa},
\nonumber
\\
&Q^{Aa} \bar{\lambda}^{Bb} = (\gamma_i)^{AB} \epsilon^{ab} D^i \phi -
i \epsilon^{AB} (\sigma_\alpha)^{ab} D^\alpha \phi,
\tag{A.9}
\\
\intertext{and}
&\bar{Q}_{Aa} G_i = (\gamma_i)_A^{\!~~B} \lambda_{Ba},
\nonumber
\\
&\bar{Q}_{Aa} V_\alpha = i (\sigma_\alpha)_a^{\!~~b} \lambda_{Ab},
\nonumber
\\
&\bar{Q}_{Aa} \bar{\phi} = 0,
\nonumber
\\
&\bar{Q}_{Aa} \bar{\lambda}^{Bb} = \hbox{$\frac{1}{4}$}
(\gamma_{ij})_A^{\!~~B} \delta_a^{\!~~b} F^{ij} -
\hbox{$\frac{1}{2}$} i (\gamma_i)_A^{\!~~B} (\sigma_\alpha)_a^{\!~~b}
F^{i \alpha} + \hbox{$\frac{1}{4}$} i \delta_A^{\!~~B}
\epsilon_{\alpha\beta\gamma} (\sigma^\gamma)_a^{\!~~b} F^{\alpha\beta} -
\delta_A^{\!~~B} \delta_a^{\!~~b} [ \bar{\phi}, \phi ],
\nonumber
\\
&\bar{Q}_{Aa} \phi = - \bar{\lambda}_{Aa},
\nonumber
\\
&\bar{Q}_{Aa} \lambda_{Bb} = - (\gamma_i)_{AB} \epsilon_{ab} D^i \bar{\phi} -
i \epsilon_{AB} (\sigma_\alpha)_{ab} D^\alpha \bar{\phi}.
\tag{A.10}
\end{align}

In order to infer from (A.8)--(A.10) the structure of the
cohomological $N_T = 3$, $D = 8$ SYM according to the
decomposition (\ref{2.9}) we have to replace ($G_i, V_\alpha$) and
$\lambda_{Aa}$ through $A_{Aa}$ and ($\chi_i, \eta_\alpha$),
respectively, in an appropriate manner. To this end, we view
$\lambda_{Aa}$ and $\bar{\lambda}_{Aa}$ in the Euclidean space as
two independent $8$--component spinors, so that they are no longer
related by hermitean conjugation. Hence, hermiticity is abandoned
--- which is not a problem here since hermiticity is primarily necessary for
the unitarity of the theory, and unitarity only makes sense for
theories in Minkowski space.

Next, we express $F_{ij}$, $F_{i \alpha}$ and $F_{\alpha\beta}$ in terms of
the field strength tensor $F_{AB ab}$ according to
\begin{align}
&F_{ij} = \hbox{$\frac{1}{4}$}
(\gamma_{ij})_{AB} \epsilon_{ab} F^{AB ab},
\nonumber
\\
&F_{i \alpha} = \hbox{$\frac{1}{4}$} (\gamma_i)_{AB} (i
\sigma_\alpha)_{ab} F^{AB ab}, \nonumber
\\
&F_{\alpha\beta} = \hbox{$\frac{1}{4}$} \epsilon_{AB}
\epsilon_{\alpha\beta\gamma}  (i \sigma^\gamma)_{ab} F^{AB ab},
\tag{A.11}
\end{align}
so that, by making use of the completeness relations
\begin{equation}
\tag{A.12} (i\sigma_\alpha)_{ab}  (i\sigma^\alpha)_{cd} =
\Omega_{abcd}, \qquad (\gamma_i)_{AB} (\gamma^i)_{CD} =
\Omega_{ABCD},
\end{equation}
where $\Omega_{abcd}$ and $\Omega_{ABCD}$ have been introduced in
(\ref{2.4}) and (\ref{2.6}), respectively, we can split $F_{AB
ab}$ into
\begin{equation}
\tag{A.13} F_{AB ab} = \hbox{$\frac{1}{4}$} (\gamma_{ij})_{AB}
\epsilon_{ab} F^{ij} + \hbox{$\frac{1}{4}$} (\gamma_i)_{AB}
(i\sigma_\alpha)_{ab} ( F^{i \alpha} - F^{\alpha i} ) +
\hbox{$\frac{1}{4}$} \epsilon_{AB} \epsilon_{\alpha\beta\gamma}
(i\sigma^\gamma)_{ab} F^{\alpha\beta},
\end{equation}
which corresponds to the $Sp(4) \otimes Sp(2)$ decomposition
$\mathbf{28} = \mathbf{(1,3)} \oplus \mathbf{(5,3)} \oplus
\mathbf{(10,1)}$ of the adjoint representation of $SO(8)$. Notice
that the relative factors in (A.11) and (A.13) are the same. The
relationships (A.11) and (A.13) immediately suggests how one has
to carry out the above mentioned replacements.

Namely, from (A.8) we deduce that the cohomological action, with
an underlying $N_T = 3$ equivariant shift symmetry, should be of
the form
\begin{align}
S^{(N_T = 3)}_{\bigr|Sp(4) \otimes Sp(2) \subset SO(8)} = \int
d^4z\, d^4\bar{z}\, {\rm tr} \Bigr\{& \hbox{$\frac{1}{4}$}
F^{ABab} F_{ABab} + 2 D^{Aa} \bar{\phi} D_{Aa} \phi - 2 [
\bar{\phi}, \phi ]^2 \nonumber
\\
& + 2 \bar{\lambda}^{Aa} (\gamma^i)_A^{\!~~B} D_{Ba} \chi_i +
2 i \bar{\lambda}^{Aa} (\sigma^\alpha)_a^{\!~~b} D_{Ab} \eta_\alpha
\nonumber
\\
& - 2 \phi \{ \chi^i, \chi_i \} - 2 \phi \{ \eta^\alpha, \eta_\alpha \} -
2 \bar{\phi} \{ \bar{\lambda}^{Aa}, \bar{\lambda}_{Aa} \} \Bigr\}.
\tag{A.14}
\end{align}

Moreover, splitting $Q_{Aa}$ into $(Q_i, Q_\alpha)$ from (A.9) and (A.10)
we deduce that the supersymmetry transformations generated by $Q_i$,
$Q_\alpha$ and $\bar{Q}_{Aa}$ should be taken as
\begin{align}
&Q_\alpha A_{Aa} = i (\sigma_\alpha)_a^{\!~~b} \bar{\lambda}_{Ab},
\nonumber
\\
&Q_\alpha \bar{\lambda}_{Aa} = i (\sigma_\alpha)_a^{\!~~b} D_{Ab} \phi,
\nonumber
\\
&Q_\alpha \phi = 0,
\nonumber
\\
&Q_\alpha \bar{\phi} = \eta_\alpha,
\nonumber
\\
&Q_\alpha \eta_\beta = - \hbox{$\frac{1}{4}$} i \epsilon_{\alpha\beta\gamma}
(\sigma^\gamma)^{ab} \epsilon_{AB} F^{AB}_{\!~~~~ab} +
\delta_{\alpha\beta} [ \phi, \bar{\phi} ],
\nonumber
\\
&Q_\alpha \chi_i = \hbox{$\frac{1}{4}$} i (\sigma_\alpha)^{ab}
(\gamma_i)_{AB} F^{AB}_{\!~~~~ab},
\tag{A.15}
\\
\nonumber
\\
&Q_i A_{Aa} = - (\gamma_i)_A^{\!~~B} \bar{\lambda}_{Ba},
\nonumber
\\
&Q_i \bar{\lambda}_{Aa} = (\gamma_i)_A^{\!~~B} D_{Ba} \phi,
\nonumber
\\
&Q_i \phi = 0,
\nonumber
\\
&Q_i \bar{\phi} = \chi_i,
\nonumber
\\
&Q_i \chi_j = - \hbox{$\frac{1}{4}$} (\gamma_{ij})_{AB} \epsilon^{ab}
F^{AB}_{\!~~~~ab} + \delta_{ij} [ \phi, \bar{\phi} ],
\nonumber
\\
&Q_i \eta_\alpha = - \hbox{$\frac{1}{4}$} i (\gamma_i)_{AB}
(\sigma_\alpha)^{ab} F^{AB}_{\!~~~~ab},
\tag{A.16}
\\
\intertext{and}
&\bar{Q}_{Aa} A_{Bb} = i \epsilon_{AB} (\sigma_\alpha)_{ab} \eta^\alpha +
(\gamma^i)_{AB} \epsilon_{ab} \chi_i,
\nonumber
\\
&\bar{Q}_{Aa} \bar{\lambda}_{Bb} = F_{ABab} + \hbox{$\frac{1}{4}$}
(\gamma_i)_{AB} (\sigma_\alpha)_{ab} (\gamma^i)_{CD} (\sigma^\alpha)_{cd}
F^{CDcd} - \epsilon_{AB} \epsilon_{ab} [ \bar{\phi}, \phi ],
\nonumber
\\
&\bar{Q}_{Aa} \phi = - \bar{\lambda}_{Aa},
\nonumber
\\
&\bar{Q}_{Aa} \bar{\phi} = 0,
\nonumber
\\
&\bar{Q}_{Aa} \eta_\alpha = - i (\sigma_\alpha)_a^{\!~~b}
D_{Ab} \bar{\phi},
\nonumber
\\
&\bar{Q}_{Aa} \chi_i = - (\gamma_i)_A^{\!~~B} D_{Ba} \bar{\phi},
\tag{A.17}
\end{align}
respectively. After a straightforward, but tedious calculation one proves
that, in fact, the above transformations leave the action (A.14) invariant.

Finally, we rename $\bar{\lambda}_{Aa} = \psi_{Aa}$ and introduce
the fields $\chi_{AB}$ and $\eta_{ab}$ via $\chi_{AB} =
(\gamma_i)_{AB} \chi^i$ and $\eta_{ab} = i (\sigma_\alpha)_{ab}
\eta^\alpha$, respectively. Similarly, instead of $Q_i$ and
$Q_\alpha$, we introduce the supercharges $Q_{AB}$ and $Q_{ab}$
via $Q_{AB} = (\gamma_i)_{AB} Q^i$ and $Q_{ab} = i
(\sigma_\alpha)_{ab} Q^\alpha$, respectively.

In this manner, by virtue of (A.12), from (A.14) and
(A.15)--(A.17) we arrive exactly at the action (\ref{2.10})
together with the supersymmetry transformations
(\ref{2.11})--(\ref{2.13}).
\bigskip

%%%%%%%%%%%%%%%%%%%%%%%%%%%%%%%%%%%%%%%%%%%%%%%%%%%%%%%%%%%%%%%%%%%%%%%%%%%%
\noindent {\large{\bf Acknowledgement}} We are grateful to M. Blau
for an interesting and useful discussion.
%%%%%%%%%%%%%%%%%%%%%%%%%%%%%%%%%%%%%%%%%%%%%%%%%%%%%%%%%%%%%%%%%%%%%%%%%%%%

\end{document}